\documentclass[journal]{vgtc}                

\ifpdf
  \pdfoutput=1\relax                   
  \usepackage{graphicx}                
  \DeclareGraphicsExtensions{.pdf,.png,.jpg,.jpeg} 
\else
  \usepackage{graphicx}                
  \DeclareGraphicsExtensions{.eps}     
\fi%

\graphicspath{{figures/}{pictures/}{images/}{./}} 

\usepackage{microtype}                 
\PassOptionsToPackage{warn}{textcomp}  
\usepackage{textcomp}                  
\usepackage{mathptmx}                  
\usepackage{times}                     
\usepackage{cite}                      
\usepackage{tabu}                      
\usepackage{booktabs}                  
\usepackage{pbox}

\usepackage{color}

\RequirePackage{etoolbox}
\pretocmd\PackageWarning{%
    \edef\pkgname{#1}\edef\hyperrefname{hyperref}%
    \ifx\pkgname\hyperrefname
        \expandafter\gobblethree
    \fi
}{}{\undefined}
\newcommand*{\gobblethree}[3]{}

\definecolor{darkred}{rgb}{0.7,0.1,0.1}
\definecolor{darkblue}{rgb}{0.1,0.1,0.7}
\definecolor{darkgreen}{rgb}{0.1,0.7,0.1}

\newcommand{\head}[1] {}

\newcommand{\rev}[1] {{#1}}

\onlineid{0}

\vgtccategory{Research}
\vgtcpapertype{application/design study}

\title{InCorr: Interactive Data-Driven Correlation Panels\\ for Digital Outcrop Analysis}

\author{Thomas Ortner, Andreas Walch, Rebecca Nowak, Robert Barnes, Thomas H\"ollt, and M. Eduard Gr\"oller.}
\authorfooter{
\item
 Thomas Ortner, Andreas Walch, and Rebecca Nowak are with VRVis Zentrum für Virtual Reality und Visualisierung Forschungs-GmbH.\\E-mail:~\{ortner$|$walch$|$rnowak\}@vrvis.at.
\item
 Robert Barnes is with Imperial College London.\\E-mail:~robert.barnes@imperial.ac.uk.
\item
Thomas H\"ollt is with TU Delft.\\E-mail:~T.Hollt-1@tudelft.nl.
\item
M. Eduard Gr\"oller is with Technische Universit\"at Wien.\\E-mail:~meister@cg.tuwien.ac.at.
}

\shortauthortitle{Ortner \MakeLowercase{\textit{et al.}}: Interactive Correlation Panels}

\abstract{
Geological analysis of 3D Digital Outcrop Models (DOMs) for reconstruction of ancient habitable environments is a key aspect of the upcoming ESA ExoMars 2022 Rosalind Franklin Rover and the NASA 2020 Rover Perseverance missions in seeking signs of past life on Mars. Geologists measure and interpret 3D DOMs, create sedimentary logs and combine them in `correlation panels' to map the extents of key geological horizons, and build a stratigraphic model to understand their position in the ancient landscape. Currently, the creation of correlation panels is completely manual and therefore time-consuming, and inflexible. With InCorr we present a visualization solution that encompasses a 3D \rev{logging} tool and an interactive data-driven correlation panel that evolves with the stratigraphic analysis. For the creation of InCorr we closely cooperated with leading planetary geologists in the form of a design study. We verify our results by recreating an existing correlation analysis with InCorr and validate our correlation panel against a manually created illustration. Further, we conducted a user-study with a wider circle of geologists. Our evaluation shows that InCorr efficiently supports the domain experts in tackling their research questions and that it has the potential to significantly impact how geologists work with digital outcrop representations in general.
} 


\teaser{
  \centering 
  \includegraphics[width=0.99\linewidth, trim=5 560 5 5,clip=true]{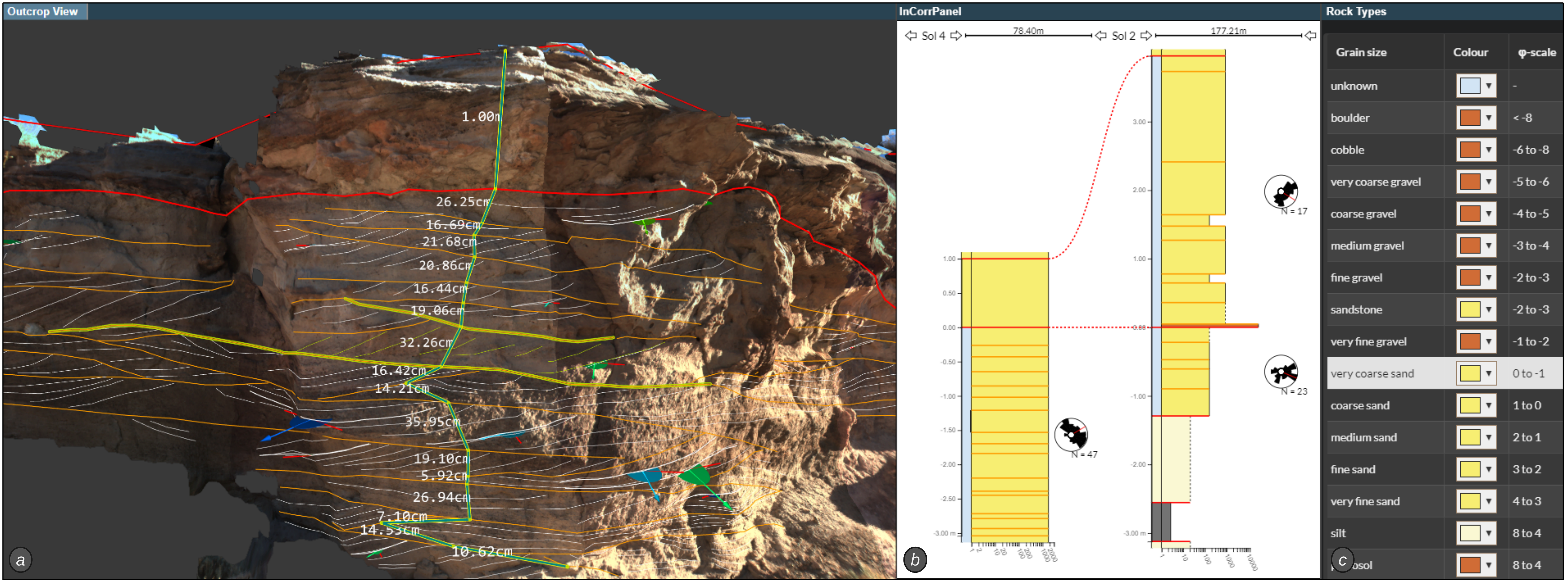}
  \caption{System overview of InCorr: (a) Outcrop View, showing digital outcrop model annotated in 3D by a geologist and vertical 3D logging tool indicating layer thicknesses. (b) InCorrPanel, showing logs for two outcrops created directly from annotation data mimicking manual illustrations. (c) GUI to assign rock types to rock layers.} \label{fig:teaser}
}

\vgtcinsertpkg

\begin{document}


\firstsection{Introduction}

\maketitle

\begin{figure*}[t]
 \centering
	\includegraphics [width=0.95\linewidth,trim=20 570 330 20,clip=true] {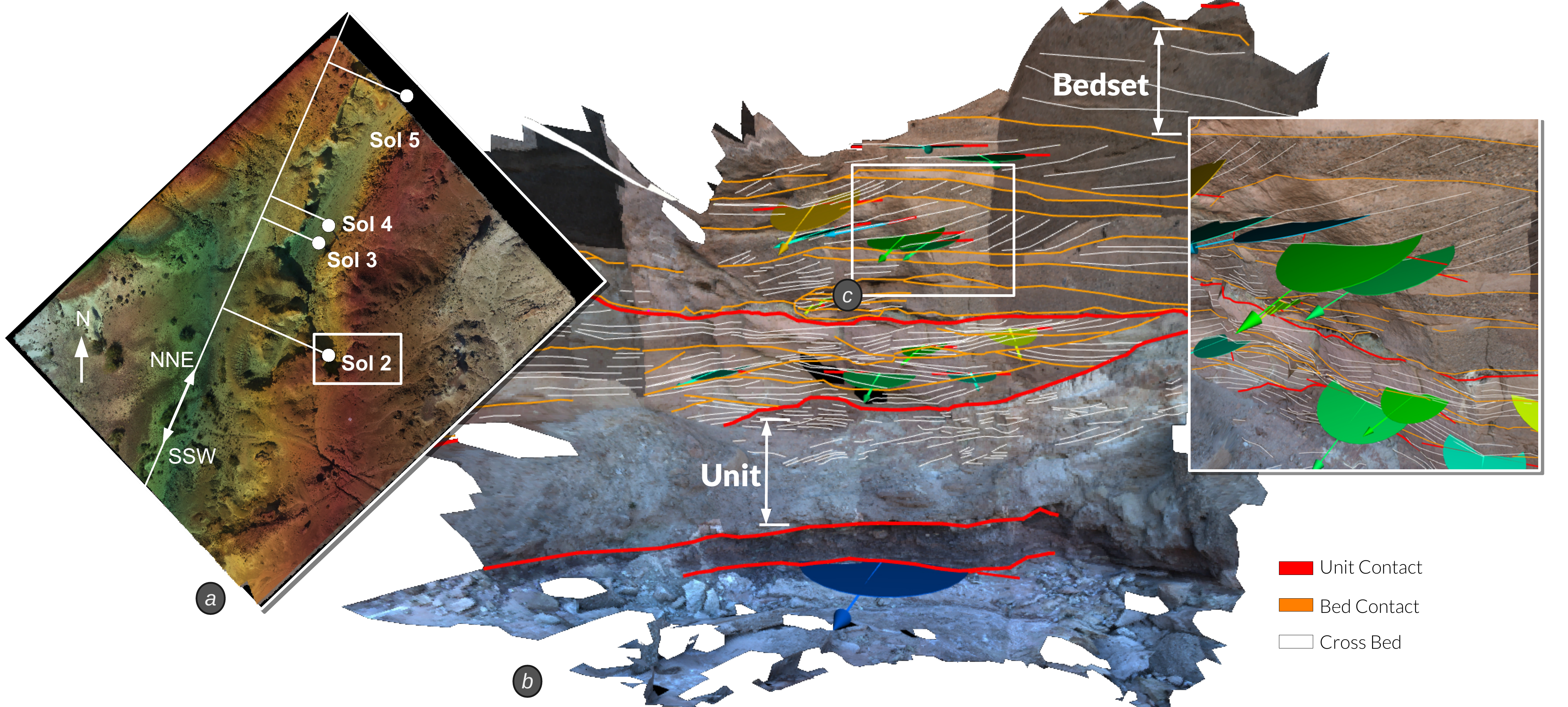}
 \caption{(a) Map of the Hanksville-Burpee Dinosaur Quarry campaign with its four outcrops along the canyon. (b) 3D triangulated mesh as digital outcrop model (DOM) of Sol2 with interpretation showing bedsets and units and their respective contacts. (c) Cross bed measurements, which determine a layer's deposition direction indicated by dip-and-strike disks. \vspace{-0.3cm}}
 \label{fig:outcrop}
\end{figure*}

\head {Geology} Geological analysis of image data collected by stereo camera systems on Mars rovers has proven to be a valuable tool in reconstructing ancient environments on Mars. This is a key aspect of the upcoming ESA ExoMars 2022 \emph{Rosalind Franklin} Rover and the NASA 2022 Rover \emph{Perseverance} missions \cite{esalife,nasalife}. A shared aim of both of these missions is to drill and sample rocks, which were deposited in ancient environments that scientists deem may have been habitable. This is determined largely by the geological characteristics. Within geology, the field of sedimentology is concerned with the analysis of textures and internal fabrics of rocks formed by deposition and movement of sand, silt, and clay sediment by wind, water, or ice. Burial and exposure to high heat, pressure and fluid circulation leads to the formation of sedimentary rocks.
Stratigraphy is a field concerned with documenting and interpreting the vertical distribution of different textures and sedimentary structures in order to reconstruct the evolution of the environments, which formed the sedimentary layers, or \emph{strata}. There is strong scientific evidence, that there were rivers and lakes active at the surface in the distant past \cite{Grotzinger2015}. Assuming analogous processes on Mars as on Earth, lake and river deposits are most promising for scientists to discover biosignatures. Therefore, stratigraphic analysis and correlation of observations between distant locations is essential.

\head{DOM} The main goal of geologists is to combine their observations to build a geological model, which encompasses the temporal evolution of environmental processes of a region. They build such a model from meticulously annotating and measuring textural features in a number of \emph{outcrops}, i.e. rock faces exposing strata as illustrated in \autoref{fig:outcrop}, and determining their  relative age-relationships.
It is generally accepted that younger sediments are deposited on older sediments, therefore changes in rock characteristics are a record of its past. Correlation of these observations across large areas allows for regional evolutionary models to be built. Traditionally, geologists make measurements `in the field' using hands-on tools, such as a compass clinometer to measure surface orientation. The development of affordable remote sensing solutions has prompted a sharp rise in geological analyses of \emph{digital outcrop models (DOM)}. 
DOMs present the geologists with 3D triangulated, and often photographically textured, surfaces of rock outcrops. Application of these techniques to processed stereo-camera image data collected by rovers on Mars can be used to greatly enhance our understanding of the evolution of the planet and drive future robotic exploration missions.

\head{Correlation Panels} Robust and efficient interpretations and measurements can be collected from DOMs in PRo3D \cite{barnes2018geological}. These data are the foundation for creating a regional geological model.  Outcrops can be understood as sampling locations, partially exposing a succession of strata, which in fact may extend over hundred thousands of square kilometers underneath a planet's surface. To build a regional geological model, geologists look for occurrences of the same stratum in multiple outcrops. First, they characterize the succession of exposed strata from an outcrop by creating a \textit{geological log}, as shown in \autoref{fig:teaser}a. After repeating this for each outcrop, they combine all logs into a \emph{correlation panel}, and connect matching strata to form \emph{correlations}, as indicated by the red lines in \autoref{fig:teaser}b. Correlating strata across multiple outcrops that cover a large area leads to a reliable characterization of the distribution of important geological units.

\subsection{Challenges} \label{sec:intro}
In a long lasting cooperation with planetary geologists, who are engaged in the geological survey of Mars as part of the scientific working groups of ESA and NASA missions, we identified the following challenges that come with remote geological analysis.

At the moment, DOM visualization software is tailored towards the \emph{interpretation} of outcrops, i.e. annotating and measuring strata. For further analysis, geologists typically rely on the export of measured values into tools for statistical analysis and graph plotting, while they create correlation panels manually based on the obtained values.
Geologists often use vector drawing software or they might draw the panels completely by hand. In any case, the creation of correlation panels is very time-consuming and therefore it is typically left to the end of the workflow.
However, only then potential weaknesses in the interpretation data become apparent, such as, insufficient detail in the interpretation of a particular outcrop. This requires a tedious revisit of the interpretation stage and forces geologists to reorganize and largely redraw their panels. Further, the separation between digital tools enforces a separation of interpreting and correlating, which would not be the case in the workflow of the traditional field work. Ultimately, this separation disconnects correlation panel illustrations from the underlying annotations and measurements, which inhibits traceability, reproducibility, and reusability of analysis results throughout the geological science community.

\subsection{Design Goals} \label{sec:goals}
Based on these observations we came to the hypothesis that the workflow of remote geological analysis would significantly benefit from an interactive and data-driven correlation panel, which achieves the following goals: \vspace{-0.2cm}

\begin{itemize}
  \setlength\itemsep{0em}
	\item \textbf{G1} The panel should be completely data-driven and consequently evolve with the analysis with minimal effort.
	\item \textbf{G2} Allow geologists to relate visual representations in the correlation panel to the data they were created from.
	\item \textbf{G3} Such a panel should mimic manually \rev{illustrated} correlation panels including stylistic freedom, but without overwhelming customization options.
	\item \textbf{G4} Using this panel needs to integrate well with the geologists' tool chain and workflow, otherwise it will not be used frequently.	
	\item \textbf{G5} The panel should intuitively put 2D geological logs in context to each other and the 3D outcrop interpretation.	
\end{itemize}
\vspace{-0.2cm}

\begin{figure*}[t!]
 \centering
	\includegraphics [width=0.95\linewidth,trim= 33 622 310 20,clip=true] {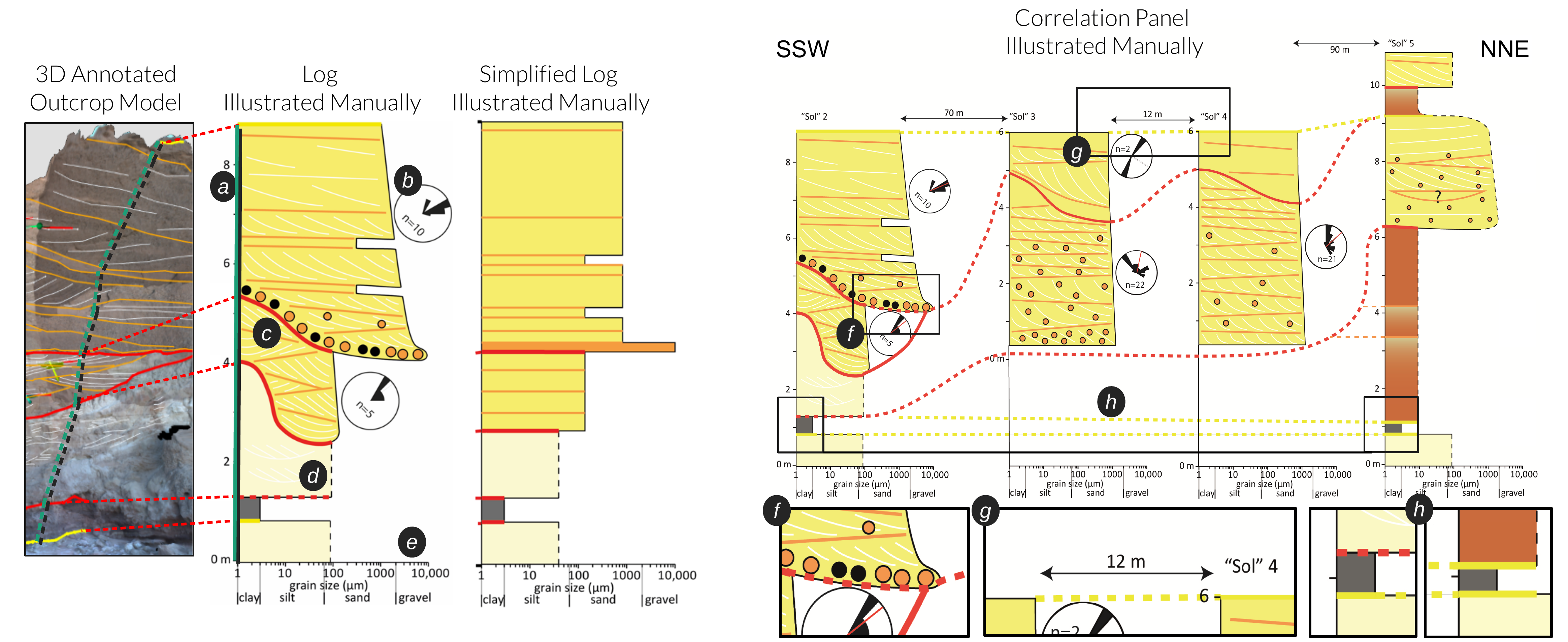}
 \caption{ \rev{(left) Annotated 3D outcrop model, followed by a sophisticated and a simplified log representation, both illustrated manually with a vector drawing tool.} (a) y-axis encodes true thickness of strata. (b) Rose diagrams show the distribution of dipping orientations. (c) Individual styles and glyphs may convey rock characteristics. (d) Dashed lines convey uncertainty. (e) x-axis encodes grain sizes logarithmically, directly relating to rock types. (right) Logs arranged in a \rev{manually illustrated} correlation panel (f) with converging contacts. (g) Spatial distance between logs along geographic direction. (h) Fossil layer used as leveling horizon. \vspace{-0.3cm}}
 \label{fig:log}
\end{figure*}

\subsection{Contributions}

As the primary contribution we present the visualization solution \mbox{\emph{InCorr}}, short for \textit{In}teractive data-driven \textit{Corr}elations. Based on the long lasting cooperation with leading scientists in the field of planetary geology, we conducted a design study to create a 3D geological logging tool (\autoref{fig:teaser}a) and the \textit{InCorrPanel}, an interactive data-driven correlation panel (\autoref{fig:teaser}b).
Both components are integrated into PRo3D \cite{barnes2018geological}, a tool for the geological interpretation of DOMs.
We verify the applicability of InCorr through a use case from a terrestrial campaign near Hanksville, Utah, USA and we validate a correlation panel generated with InCorr against a manually \rev{illustrated} one based on the same interpretation data (\autoref{fig:log}).
\rev{We further conducted a hands-on design validation to collect feedback from a broader range of geologists}. Another contribution is the introduction of interactive correlation panels to the domain of remote geology analysis, which fosters traceability, reproducibility, and communicability of an otherwise static illustration.

\section{\rev{Previous Work}}

The Petrel~\cite{schlumbergerpetrel} software package for oil and gas exploration offers some outcrop measurement capabilities and correlation panels.
However, as the software was originally intended for the analysis of seismic data in connection with drill shafts, the logs present in these panels are created from drilling wells and have a different visual encoding. Therefore, they are not suitable for performing outcrop-based correlation analysis. 
General purpose GIS or 3D visualization tools, such as ArcGIS~\cite{arcgis} or Cloud Compare~\cite{cloudcompare}, are commonly found in geological publications that include DOM analysis.
These applications offer reliable measurement tools, but are neither targeted towards outcrop interpretation applications nor do they support correlation analysis.
Respective publications dealing with DOM interpretations typically describe a concatenation of data transformations \cite{Lanen2009, Sahoo2015}.
A few specialized 3D outcrop interpretation tools have recently emerged, including LIME~\cite{buckley2019lime}, VRGS~\cite{hodgetts2007integrating}, or PRo3D~\cite{pro3dpage, barnes2018geological}.
All three feature a tool set for creating annotations and performing measurements on DOMs.
Additionally, LIME and VRGS allow geologists to project \rev{manually illustrated} logs onto the 3D surface. Nesbit et al. \cite{nesbit2018} use a 3D log in the context of their stratigraphic mapping. However, this logging is not integrated into an interactive workflow and the measurements need to be translated into 2D logs manually. 

The visualization of geological phenomena has been an essential part of visualization research for decades, but mostly in the context of the analysis of seismic data for oil and mining.
Patel et~al.~\cite{patel2008seismic} introduce a tool to interpret 2D slices of seismic data from which they can pre-calculate horizon structures.
H\"{o}llt et~al.~\cite{hollt2011interactive, hollt2012seivis} present an interactive workflow for interpreting the 3D data directly by incorporating well logs retrieved from drilling. The steps of data retrieval, interpretation, well correlation and horizon extraction, and reservoir modeling~\cite{natali2013modeling} do align with the workflow of digital outcrop analysis, described in \autoref{sec:background}, however the data, methods, and challenges differ significantly.
Lidal et al.~\cite{lidal2013geological} focus on the communication of geological processes through visual stories and provide a sketch-based interface for the creation thereof.
To the best of our knowledge the field of digital outcrop analysis and the correlation of 3D outcrop interpretation data has not been explored or addressed by the visualization community.

\section{Geological analysis of Digital Outcrop Models} \label{sec:background}

\head{workflow} The basis of remote virtual outcrop geology is the acquisition of 3D data from rock outcrops. In this section we will discuss the development of outcrop observations to regional geological models. We focus on the stages relevant to our work and introduce the geological concepts that are important throughout this paper. \emph{Outcrop interpretation} (\autoref{sec:interpret}) produces contacts, which are the basis for \emph{logging} (\autoref{sec:log}, \autoref{sec:truethickness}). The result of the logging stage are geological logs, which are arranged in a correlation panel to discover and create \emph{correlations} (\autoref{sec:panel}). Correlations are used to reconstruct geological surfaces, which are then presented as 2D geological maps. \rev{The relevant geological terms introduced in this section are summed up in \autoref{tab:glossary}.}

\subsection{Outcrop Interpretation} \label{sec:interpret}
\head{interpretation} Traditionally, geologists examine multiple outcrops in the field. At each outcrop they record textural characteristics, such as grain or crystal size and shape, relative color, layer thickness, layering patterns, layer orientation, and boundary geometries. Relevant measurements, photographs, notes, and outcrop sketches, as well as sketch logs or cross sections where necessary, are used to document the geology in the field, and analyzed out of the field for publication. The same process can be applied to DOM analysis. After measuring the height and width of an outcrop, geologists investigate \emph{contacts}. A contact represents the delineation where one layer of rock, i.e. \emph{stratum}, ends and another one begins. These contacts can be numerous and are typically nested, so a stratum contains sub-strata, as shown by orange lines (\emph{bedsets}) between red lines (\emph{units}) in \autoref{fig:outcrop}b. Based on the different visible rock characteristics, geologists meticulously trace contacts by drawing polylines on the 3D surface along the discrete transitions between the deposited strata. This allows them to characterize the architecture of the strata of an outcrop hierarchically into sub-structures of arbitrary depth. Two contacts delimit a stratum that is homogeneous (to a certain extent). Geologists use different line thicknesses and colors to represent the magnitude of change between two adjacent strata. Geologists refer to this process of identifying contacts and hierarchical grouping as \emph{geological interpretation}.

\head{cross bedding, dns} \emph{Cross beds} are stratifications within a stratum, visible as thin white lines in \autoref{fig:outcrop}b, which are the preserved lee-faces of dunes or ripples, which migrated by wind or water action. The azimuth of the maximum dip direction of these cross beds indicates the original transport direction of the respective deposition medium~\cite{prothero1996introduction}.
Therefore, the interpretation of cross beds is invaluable to determine the direction of wind or water flowing in the geological past.
In the field, geologists use a compass-clinometer. In PRo3D, they trace a cross bed by picking 3D points on the DOM surface. Then a plane is fitted to these points via total-least square regression \cite{jones2016robust}. The result is a so-called \emph{dip-and-strike} measurement, where the \emph{dip} is the direction of maximum negative inclination of this plane, while the \emph{strike} is orthogonal to the dipping direction, as illustrated in \autoref{fig:outcrop}c. In the context of correlation panels, geologists are primarily interested in the geographic direction of dips. Geographical directions are quantified as azimuth in degrees, where 0\textdegree, 90\textdegree, 180\textdegree, and 270\textdegree point to north, east, south, and west respectively.

\head{grain size} The sizes of grains found in a stratum are a decisive factor, to determine its rock type and its mode of deposition, that is either by wind (aeolian) or by water (fluvial). Grain sizes range from coarse soil, such as cobble with 63-200 mm, to fine soil, such as clay with a grain size smaller than 0.002 mm. While in the field, geologists have many tools at their disposal, measuring grain sizes in DOMs is limited to grains visible on the textured mesh. For the exploration of Mars, scientists mostly have to rely on image-derived data, so they often need to infer grain sizes, rather than being able to measure them in 3D.

\begin{figure}[b!]
 \centering
	\includegraphics [width=0.6\linewidth,trim=100 110 100 60,clip=true] {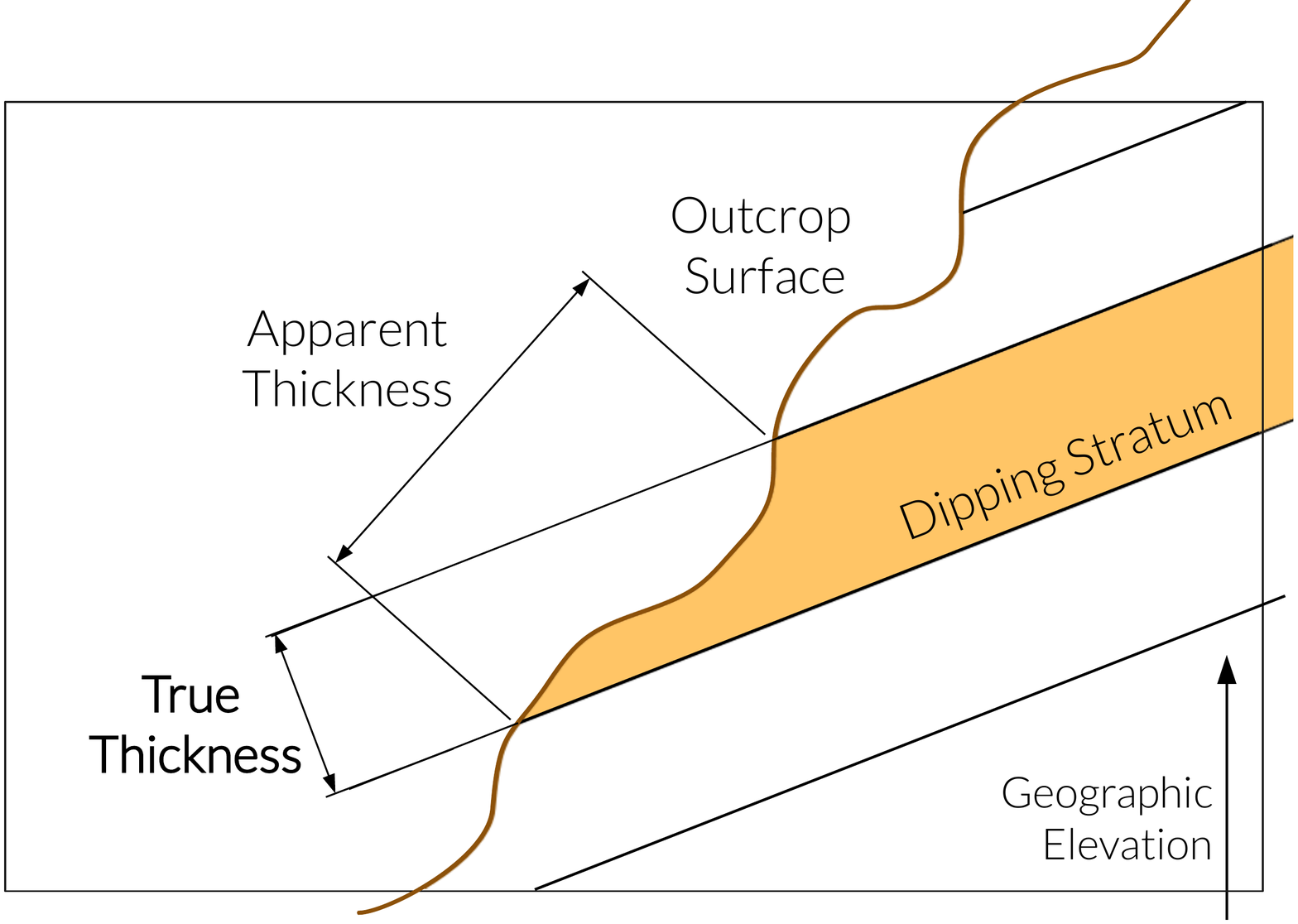}
 \caption{The apparent thickness of a stratum  observed at an outcrop is often misleading. The rock may be broken off at a slanted angle or the stratum itself may be tilted. A dip-and-strike measurement is essential to determine a stratum's orientation and compute its true thickness. \vspace{-0.3cm}}
 \label{fig:truethickness}
\end{figure}

\subsection{Geological Logs} \label{sec:log}

\head{log creation} A 3D interpretation that characterizes an outcrop may contain a plethora of measurement values and annotations. For the sake of clarity, we will focus on the ones that are essential for creating a \emph{geological log}, as depicted in \autoref{fig:log}.
A log characterizes the sequence of strata as they were deposited over time, starting with the oldest at the bottom and ending with the most recently deposited at the top. Perfectly horizontal strata are rare; geological effects and the roles of deposition and erosion in landscape evolution can produce irregular contacts. Therefore, the geologists `draw' a log over the interpretation, connecting the contacts in their vertical, i.e. chronological sequence. The difference in elevation between two contacts determines the thickness of a stratum. Currently no 3D interpretation tool does support the direct semantic connection of contacts to create a geological log. Instead, elevation values and names of contacts are exported, and then geologists manually draw the corresponding log, while cross-checking with distance measurements in the 3D visualization. The notion of \emph{true thickness} \rev{complicates this matter significantly, which we discuss in \autoref{sec:truethickness}}. \head{log encodings} A geological log characterizes an outcrop by showing the type and succession of strata in an abstracted form. As illustrated by \autoref{fig:log}a, the y-axis of the log encodes the elevation of the contacts measured in the 3D view as well as the thickness of the strata enclosed by the respective contacts. The x-axis in the log (\autoref{fig:log}e) corresponds to the grain size on a logarithmic scale. Grain size relates to the rock type of a stratum, which is also encoded in the stratum's color. To characterize the orientation of a stratum, geologists visualize the distribution of cross bed dipping-azimuths in a rose diagram. In the example in \autoref{fig:log}b, the upper unit between the yellow and the red contact `dips towards east/north/east' based on ten dip-and-strike measurements. Dashed lines are often used to convey uncertainty, for instance, concerning a contact (horizontal line) or concerning the grain-size (vertical line) shown in \autoref{fig:log}d. Geologists use additional encodings such as curved lines for contacts of varying elevation or glyphs to convey \rev{grain} distributions (\autoref{fig:log}c).

\subsection{True Thickness} \label{sec:truethickness}

\head{true thickness} For the sake of simplicity, we accepted that the difference in geographic elevation between two contacts results in the thickness of the enclosed stratum. In nature that is often not the case, which is why geologists distinguish the measured or \emph{apparent thickness} and the \emph{true thickness} of a stratum. Our previous simplification is only valid if the measured strata lie in a horizontal plane and the outcrop surface is vertical. \autoref{fig:truethickness} illustrates the discrepancies between geographic elevation, apparent thickness, and true thickness. In reality, the deposition of material originally occurs horizontally, but strata may be tilted or even folded over by a variety of geological or geomorphological phenomena.
\rev{Hence, instead of using global elevation values over all strata and logs, each apparent thickness needs to be corrected by the stratum's dipping angle. However, in stratigraphy it is an accepted simplification to use one angle per log, which still results in each log creating its own coordinate system.} Throughout this paper, when we speak of thickness, we mean true thickness, since the apparent thickness is of little relevance for geological interpretation.

\begin{table}[t!]
\caption{\rev{Summary of the most important geological terms}.}
\begin{tabular}{|l|l|}
\hline
outcrop      &   exposed rock face showing \cr & sedimentary structures \\ \hline
stratum / -a  & layer of rock bounded by two contacts  \\ \hline
contact      & discrete transition between two strata \\ \hline
dip      & inclination angle of a stratum \\ \hline
dip-and-strike & measurement to determine the orientation \cr & of a stratum                            \\ \hline
true thickness & thickness of a stratum with respect to its dip \\ \hline
cross bed    & cross stratifications within a stratum \\ \hline
unit, bedset & specific strata, where a bedset is a \cr & sub-stratum of a unit \\ \hline
\end{tabular}
\label{tab:glossary}
\vspace{-0.5cm}
\end{table}

\subsection{Correlation Panel} \label{sec:panel}

\head{creating correlations} After having created multiple logs, geologists arrange them in a correlation panel, as shown in \autoref{fig:log} on the right. The juxtaposition of logs allows geologists to identify similar strata across outcrops and to \emph{correlate} them, hence the name correlation panel. The notion of geological correlation is not related to the mathematical concept. Correlations are visualized as colored connections between contacts. A dashed pattern is used to express uncertainty if two contacts belong to the same structure. The arrangement of logs from left to right is often determined by their succession along the course of a geomorphological feature, such as a canyon or a crater rim. Rulers between the individual logs encode the distances between them (\autoref{fig:log}g). The distances also convey a degree of uncertainty and data quality, since inferring correlations between outcrops across large distances is less reliable. Due to geological faults or other geomorphological processes, different outcrops do not necessarily expose the same stratum at the same elevation or with the same thickness. Consequently, correlation lines are rarely horizontal, which is why geologists often choose a distinct contact as a leveling horizon for all outcrops. In the example shown in \autoref{fig:log}, a stratum rich of fossils has been found in first and the last outcrop, indicated by dark gray rectangles at an elevation of 1m. This stratum is missing from the other logs, because it is not exposed at these locations and potentially buried. Still, geologists are able to infer its existence and its approximate elevation.

\subsection{Hanksville-Burpee Dinosaur Quarry, UT, USA} \label{sec:hbdq}

Image data was collected at the Hanksville-Burpee Dinosaur Quarry (HBDQ), near Hanksville, Utah, U.S.A. (110$^{\circ}$47'30"W, 38$^{\circ}$27'12"N). The distinctive rocks of the Jurassic Morisson Formation have a sub-horizontal regional dip, and are therefore ideal for testing the capabilities of PRo3D in manually creating and correlating geological logs. During this campaign, the geologists digitally captured four outcrops along a canyon, as illustrated by the map in \autoref{fig:outcrop}a, named \mbox{Sol2--5}. After capturing, they reconstructed DOMs using PRoViP \cite{gao2016contemporary} and interpreted each of them in PRo3D yielding contacts and cross bed measurements.  Based on the interpretations they manually measured the true thicknesses of the strata and drew a log for each outcrop. All outcrop interpretations of this dataset do contain two specific strata, \emph{units} enclosed by unit contacts in red and \emph{bedsets} enclosed by bed contacts in orange, as illustrated in \autoref{fig:outcrop}. They arranged the individual logs in a manually illustrated panel and preliminary derived correlations. The resulting correlation panel is visible in \autoref{fig:log}. We use the DOMs, the interpretations, the logs, and the correlation panel as a running example throughout this paper and as a use case in \autoref{sec:eval}.

\begin{figure}
 \centering
	\includegraphics [width=0.95\linewidth] {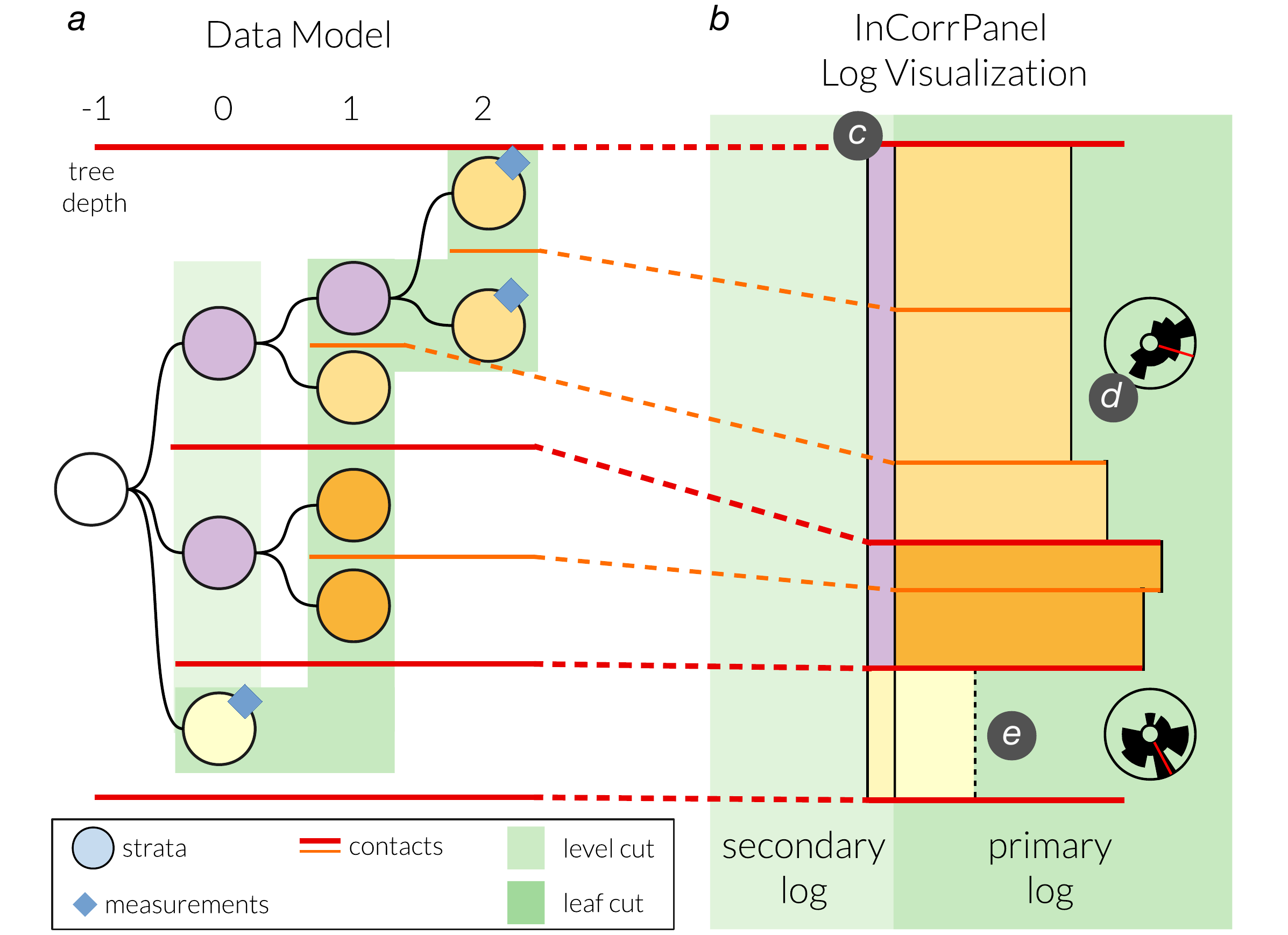}
 \caption{(a) Data model: tree of strata and their bounding contacts. (b) InCorrPanel: (c) secondary log shows a tree cut at depth 1, while the primary log shows the leaf cut. (d) Rose diagrams aggregate cross bed orientations. (e) Dashed border conveys uncertain an rock type. \vspace{-0.3cm}}
 \label{fig:logvis}
\end{figure}

\section{\rev{Design Process and Domain Abstraction}}

\rev{The results of InCorr are based on a decade-long collaboration with planetary geologists, mainly in the context of research and development of PRo3D as an interpretation tool. In a workshop following the evaluation campaign described in \autoref{sec:hbdq}, our collaborators stated the need to semi-automatically generate correlation panels from interpretations and suggested a simplified visual encoding for logs, as shown in \autoref{fig:log}. We followed a participatory design approach~\cite{janicke2020participatory} leading to a three-phase evolution of InCorr: In phase \textbf{(1)}, we were concerned with how to transform annotations into a log, researching a hierarchical data structure and the transformations necessary. This resulted in an non-interactive log prototype matching the simplified visual encoding. In phase~\textbf{(2)}, we focused on understanding the domain background and tasks involved with correlation analysis and created an interactive prototype. It is integrated with PRo3D featuring multiple logs and correlations. In phase~\textbf{(3)}, the necessity of a logging tool measuring true thickness became evident. Key aspects of this phase were interaction and visualization design, and end-to-end workflow integration. Each phase was accompanied by a week-long research stay and roughly quarterly meetings. In phase~\textbf{(3)} we shortened intervals and iterated on visualization and interaction prototypes sometimes on a daily basis, shaping InCorr through the continuous feedback provided by our collaborators.}

\subsection{Tasks} \label{sec:tasks}

With InCorr we address a set of tasks that bridges the gap between outcrop interpretation and the creation of a geological model based on logs and correlations:

\begin{itemize}
  \setlength\itemsep{-0em}
	\item \textbf{T1} Create a geological log for each outcrop based on the annotations and measurements taken.
  \item \textbf{T2} Create correlations from geological logs as the basis for a regional geological model
	\item \textbf{T3} Edit and export the correlation panel to be manipulated in other tools for further analysis or dissemination \vspace{1.0cm}
\end{itemize}

We abstract these tasks and their subtasks by following the multi-level task typology by Brehmer and Munzner \cite{Brehmer2013}. We subdivide the creation of a geological log (\textbf{T1}) into connect contacts (\textbf{T1a}), add rock type (\textbf{T1b}), and add cross beds (\textbf{T1c}). In \textbf{T1a} the geologists \texttt{identify} the relevant contacts they want to connect and then \texttt{select} them to form a log (\texttt{annotate}). For each stratum in the log they \texttt{identify} the grain size and \texttt{select} the respective rock type (\textbf{T1b}). Further, they \texttt{identify} and \texttt{select} cross bed measurements belonging to a stratum to \texttt{summarize} the orientation distribution within it (\textbf{T1c}). For \textbf{T2} the geologists \texttt{arrange} the logs from \textbf{T1} to \texttt{compare} their characteristics and find similarities (\textbf{T2a}). When found, they \texttt{select} one contact per log they want to connect and create (\texttt{annotate}) a correlation (\textbf{T2b}). When the correlation analysis is completed, geologists `tidy up' the correlation panel by vertically and horizontally \texttt{arranging} the logs and export it to be used in other tools of their workflow (\textbf{T3}).

\subsection{\rev{Data Transformation}} \label{sec:model}

\rev{\head{from contacts to strata} To bridge the gap between outcrop interpretation and 3D logging, we need to infer the hierarchical structure of strata from a set of contacts. Logging allows geologists to pick a 3D position on each contact, while the true height of these positions, i.e. elevation corrected by a dipping angle, determines the vertical sequence of the contacts. Each contact has a rank assigned by the geologist, representing the magnitude of change between two rock layers. With this information, we can derive a tree of strata. Each stratum is bounded by an upper and lower contact defining its minimum and maximum height, and thus its thickness. We achieve this by starting out with a fictive stratum with a height interval from negative to positive infinity. Our algorithm goes through the set of contacts sorted by their rank, finds the stratum with a height range matching the contact's height and splits it into two sub-strata. Performing this step for each contact yields a tree of strata in a breadth-first fashion.

\head{measurements} Each stratum can be assigned a rock type and may contain geological features, such as cross beds. To characterize the distribution of cross bed orientations within a stratum, geologists perform dip-and-strike measurements. 
Such measurements can then be assigned to leaf strata. This allows us to use simple tree traversals to aggregate measurements for arbitrary strata.
As an example, we represent the log illustration shown in \autoref{fig:log} using our data model in \autoref{fig:logvis}a. The whole outcrop, represented as root stratum, contains units (sub-strata) defined by the red contacts. The upper two units contain sets defined by orange contacts, which do not have descendant strata, but contain cross beds (white lines). Dipping-azimuth values of the cross beds are aggregated at a unit level and visualized as rose diagrams. \head{correlations} We define a correlation as a set of contacts connecting them between logs. Correlations are often uncertain in parts, which means that the connection between two contacts is either certain or uncertain, typically represented as solid or dashed line, respectively. }

\begin{figure*}[]
 \centering
	\includegraphics [width=0.95\linewidth,trim=32 540 75 38,clip=true] {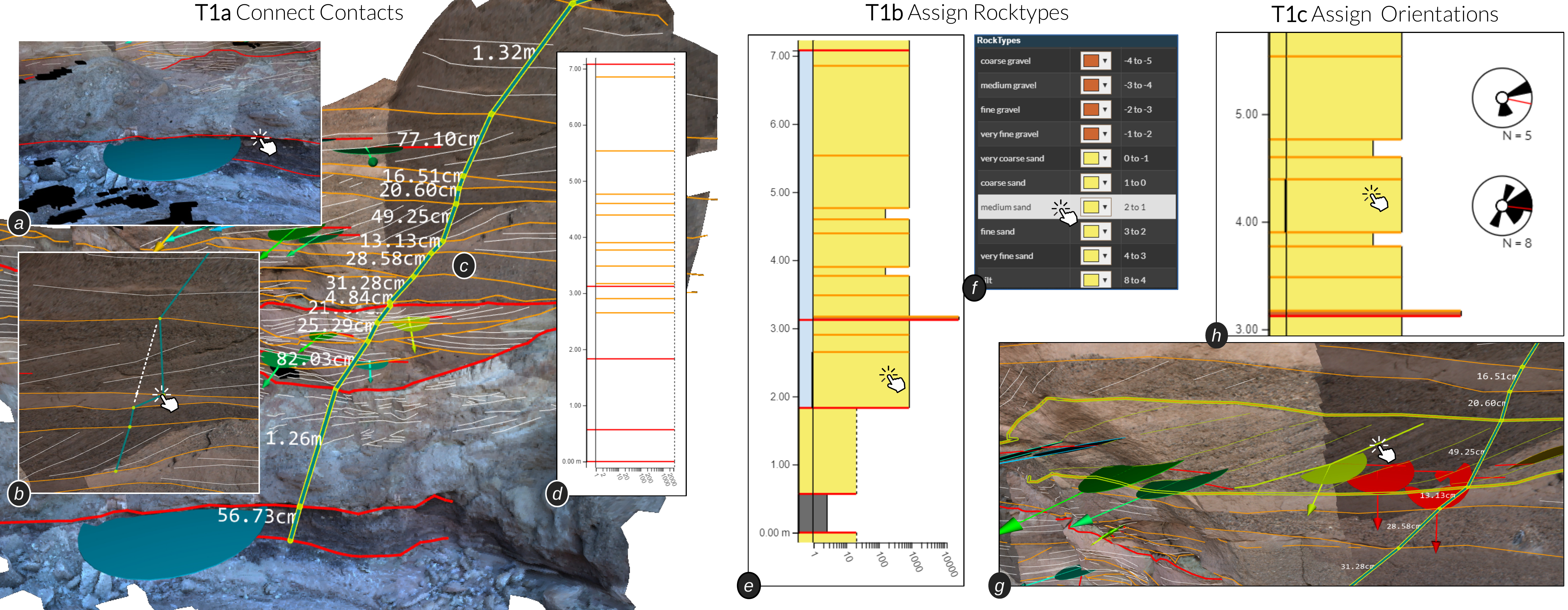}
 \caption{To create a 3D log, users (a) first pick a reference plane, then they (b) connect the contacts to form (c) a 3D log. Users then assign rock types to the (d) empty log via the (f) rock types list, resulting in (e) a log encoding rock categories and grainsizes. To finish the log, they (g) assign cross bed measurements, which are aggregated and encoded as (h) rose diagrams. \vspace{-0.3cm}}
 \label{fig:interaction}
\end{figure*}

\section{InCorr} 

\head{System Overview} InCorr consists of three components that we integrated with PRo3D: (1) the \emph{InCorrPanel}, a 2D interactive correlation panel, that offers geologists an evolving summary of their geological analysis and that is easy to keep in sync with annotations and measurements, (2) a \textit{logging tool}, that allows them to intuitively connect contacts in the Outcrop View provided by PRo3D, and (3) a \emph{list view} for assigning rock types to strata in the InCorrPanel.
In the following, we discuss the visual encodings in \autoref{sec:visual_encodings} and the interaction design in \autoref{sec:interaction} that we use for these components.

\subsection{Visual Encodings} \label{sec:visual_encodings}

\head{static panel} Geological logs are a visual abstraction of complex spatial relationships of 3D phenomena. Showing them in a correlation panel does provide geologists with a concise overview of large-scale geological analyses. Thus, static correlation panels already facilitate \textbf{T2} by enabling a visual comparison of outcrops without inspecting their spatial representations. To exploit these properties, the design of the InCorrPanel needs to closely resemble \rev{manually illustrated} logs and correlation panels, without overwhelming users with customization options~(\textbf{G3}).  

There is no universally agreed upon standard for creating geological logs. Every geologist has their own specific style. For instance, the geological logs in our running example (shown in \autoref{fig:log}) can be considered a rather sophisticated variant. Therefore, we did not conduct a formal design space analysis, but let our collaborators decide on the appropriate visual representations.
\rev{A key topic of design phase \textbf{(2)}} was to derive which aspects need standardization and which aspects need to be left to the individual geologist.
As there currently is no tool for creating correlation panels directly from outcrop interpretations, our collaborators were eager to participate in these sessions.
Hayes et~al.~\cite{Hayes2011}, Van Lanen et~al.~\cite{Lanen2009}, or Hampson et~al.~\cite{Hampson2011} present examples of the variety of correlation panels published in geological journals.
In the following we present our design decisions for InCorr starting with the visual encodings for displaying a single log.

\subsubsection{Single Log} \label{sec:singlelog}

\head{visual encoding} The visual representation of a single log in the InCorrPanel is illustrated in \autoref{fig:logvis}b.
The x-axis encodes the grain size on a logarithmic scale, while the y-axis encodes the order and true thickness between contacts.
To calculate the true thickness, a log requires a reference plane in 3D. For each contact we compute the height above the reference plane and map it directly to a position on the y-axis.
For each stratum, we draw a filled rectangle, ranging on the y-axis from the elevation of its lower contact to the elevation of its upper contact.
The stratum's grain size, which relates to the associated rock type, is encoded redundantly by the width of the rectangle and its color.
Additionally, we draw the contacts as horizontal lines at their elevation matching their thickness and color in the Outcrop View (\textbf{G2}). To indicate if geologists are uncertain with the rock type they assigned, we draw the right border of the rectangle as a dashed line \autoref{fig:logvis}e.
This is especially relevant in Martian use cases, where the tools for observation are mostly visual and the better part of the observed strata are of finer grains.
We chose the simplified rectangular log over the curved contact representations and curved polygons for several reasons. One could derive the variation in elevation by picking multiple points per log, but (1) there is no information in a contact that would describe the variation of a stratum along the x-axis;
(2) it becomes difficult to read strata thicknesses directly from the log;
and (3) the encoding is rather specific and rarely used in manual illustrations.
Ultimately, as we create the plot as an SVG image it can be directly edited in a vector image tool after export, to add custom illustration styles (\textbf{G3}, \textbf{G4}).

\head{two columns} When embedding an outcrop analysis into a bigger context, deep hierarchies of strata may emerge. To manage such hierarchies, our collaborators suggested to add a second, abstracted log representation that emphasizes the affiliation of sub-strata to their respective containing strata. We denote this abstracted log as \emph{secondary} log and show it to the left of the \emph{primary} log (\autoref{fig:logvis}c).
The primary log always shows the full detail of strata (leaf nodes of the data model), whereas the secondary log shows a single, user specified level (here level 0) of the hierarchy indicated in the data model (\autoref{fig:logvis}a).
This two-column representation of logs resembles a specialized icicle plot \cite{schulz2011treevis}, where all nodes except the leaf nodes and the selected  level are removed from the hierarchy.
A similar double log representation can, for instance, be found in the manually \rev{illustrated} correlation panel by Hampson et al. \cite{Hampson2011} (\textbf{G3}).

\head{dip aggregation} The selected depth in the secondary log also governs the granularity of how we aggregate measurement values. Supporting \textbf{T2}, the \mbox{InCorrPanel} summarizes the distribution of dip-azimuths as rose diagrams (\autoref{fig:logvis}d). 
We aggregate the orientations into 24 15$^{\circ}$ angular bins and encode the frequency into their area, as suggested by Sanderson and Peacock \cite{sanderson2019making}. This allows the geologists to infer the major dipping-azimuths of a stratum without missing low frequency outliers. We further compute the mean angle (via polar coordinates) and encode it as a red line. According to the geologists, this representation allows them to quickly judge if the distribution follows one general direction.

\subsubsection{Multiple Logs and Correlations} \label{sec:multilog} 

\begin{figure*}[]
 \centering
	\includegraphics [width=0.94\linewidth,trim=100 950 770 30,clip=true] {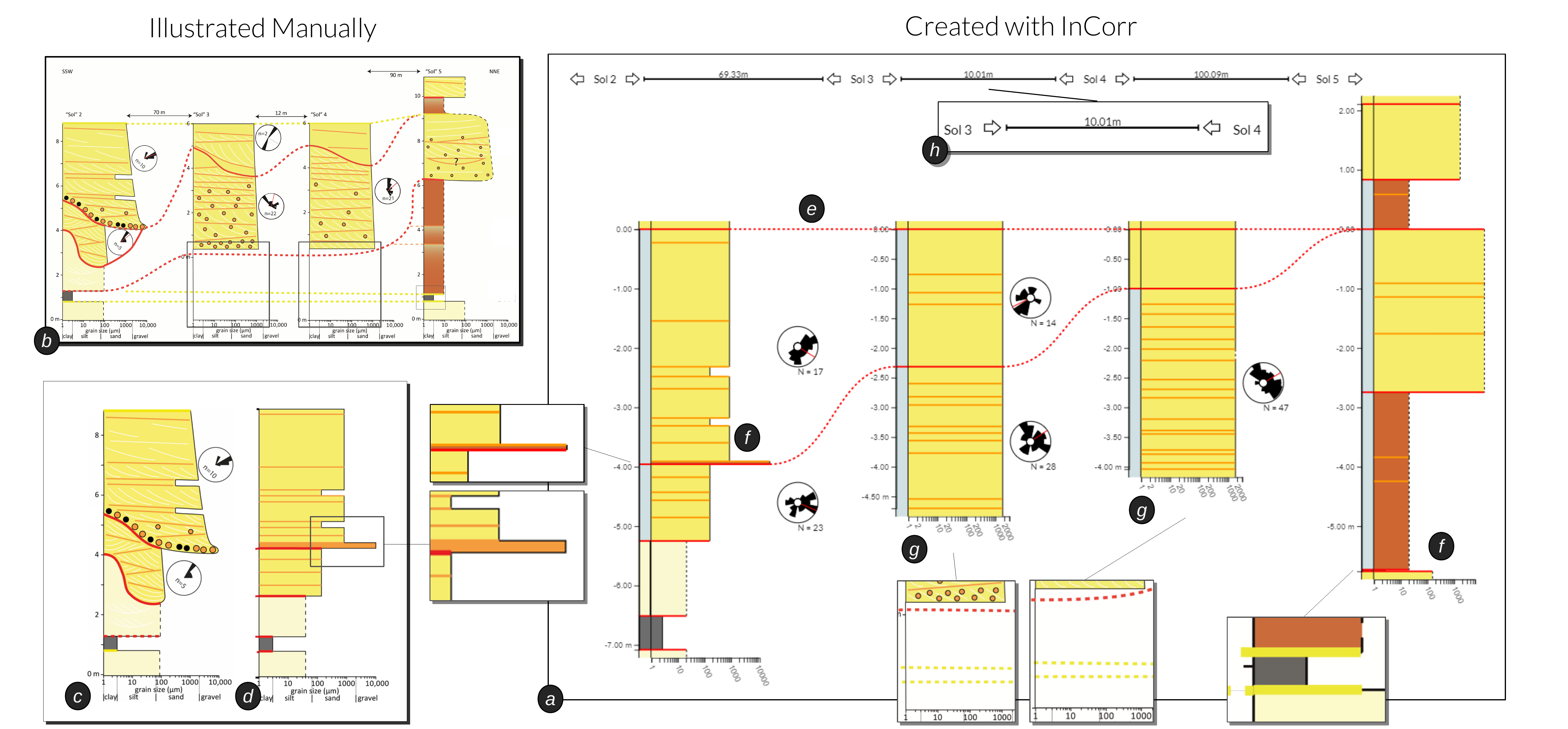}
 \caption{(a) Correlation panel created with InCorr based on outcrop interpretation data with (b,c,d) manual illustrations for comparison. (e) All logs are aligned to a common baseline correlation at the top. (f) No vertical exaggeration of strata, (g) no indication of buried regions, and no visible connection of non-adjacent logs. (h) Showing spatial distance between logs. \vspace{-0.3cm}}
 \label{fig:incorrpanel}
\end{figure*}

\head{log order} In published correlation panels, the logs are carefully ordered to convey the results in the best possible way. In the case of the HBDQ analysis, the logs are sorted along the geographic direction from SSW to NNE, as the result of a projection from longitude and latitude onto a line, as illustrated in \autoref{fig:outcrop}a.
We did consider ordering the logs in the InCorrPanel automatically along a fitted line or let the geologists pick a geographic direction. Ultimately, the log order is very context sensitive to the location of the outcrops and domain experts will want to order them manually. Each log is a concise summary of an outcrop interpretation, and a correlation panel supports geologists in identifying strata with similar characteristics. The InCorrPanel allows them to juxtapose logs arbitrarily in an interactive fashion. This enables them to quickly compare rock types and cross bed orientations during the analysis in order to correlate contacts (\textbf{T2}). This is similar to rearranging the dimensions of a parallel coordinates plot to investigate relationships. Further, we compute the distances between adjacent logs and display them as rulers between the log names, as seen in \autoref{fig:incorrpanel}h.

\head{vertical alignment} After determining the horizontal arrangement, we are left with the vertical positioning of each log in the InCorrPanel. In discussion with our collaborators, we realized that there is no single layout method for vertically positioning logs. Further, the most suitable layout is likely to change throughout the analysis. We agreed on a two-fold approach, which provides standardization as well as flexibility (\textbf{G1}, \textbf{G3}). We started out rendering all logs, their contacts, and strata in a common coordinate system based on the geographic elevation of the individual elements. Since this approach does not accommodate for true thickness we use the geographic elevation of a log's reference plane as an anchor point as discussed in \autoref{sec:singlelog}. When users find correlating strata they often use the resulting surface as baseline for vertical alignment. In the best case, every log, i.e. every outcrop, exhibits a part of the same rock stratum. Using this stratum as the baseline offsets each log vertically in such a way that all log nodes of this surface are at the same y-position in the log. In that case, the correlation lines associated with the upper contact of the stratum become horizontal. In the example in \autoref{fig:log}h, only two outcrops exhibit the baseline stratum. The dashed yellow lines in the correlation panel indicate that this surface is also present at the locations of the other two outcrops, but not exposed. 

\subsection{Interaction} \label{sec:interaction}

In this section we discuss the interactions necessary to achieve the domain tasks in the sequence described in \autoref{sec:tasks}, i.e. creating logs (\textbf{T1}), creating correlations (\textbf{T2}), and editing the resulting panel for export to other tools (\textbf{T3}). \autoref{fig:interaction} provides an overview of the succession of interactions for creating a log (\textbf{T1}), starting with the creation of a single log based on 3D contact lines from an outcrop interpretation~(\textbf{T1a}). 

\textbf{T1a Connect Contacts to Create Strata:} Before users can start connecting contacts to form a log, they need to specify a reference plane to enable true thickness computation. They choose a contact which they suspect to have a suitable orientation to which we fit a plane using 3D total-least squares regression. The fitted plane is then visualized as a colored disc, where the color encodes the dipping angle of the plane (\autoref{fig:interaction}a). Geologists can try out different contacts and inspect the respective planes or confirm the found plane. Then users pick a point on each contact they want to add to their log. Due to their implicit sequence, the selection of contacts need not to happen in order. For each picked point we compute its height above the reference plane and sort the set of points by their height over the plane. This allows us to draw 3D line segments as a pairwise combination of the elements of the sorted list (\autoref{fig:interaction}c). Further, a point picked for a contact can be modified or removed and the log polyline changes accordingly, giving immediate visual feedback to the user (\autoref{fig:interaction}b). Users can edit existing logs in the same way to keep their analysis up-to-date with new contacts~(\textbf{G1}).

\textbf{T1b Assign Rock Types to Strata:} The strata of a newly created log do not have rock types associated with them yet, indicated by their white color (\autoref{fig:interaction}d). To assign a rock type to a stratum, users first select the respective stratum in the primary log and then they pick a rock type from the rock types list (\autoref{fig:interaction}f). The color and the width of the respective rectangle changes accordingly. To indicate how sure geologists are with their choice of rock type they can add an uncertainty state shown as a dashed or solid right border encoding uncertainty or certainty, respectively. A log fully annotated with rock types, including uncertainty states, is shown in \autoref{fig:interaction}e (\textbf{G3}). When selecting a stratum in the InCorrPanel, we also highlight the bordering contacts in the Outcrop View (\autoref{fig:interaction}g), allowing geologists to identify the respective area of the outcrop (\textbf{G5}).

\textbf{T1c Assign Cross Beds to Strata:} Analogously to the previous task, users first select a stratum in the primary log in the \mbox{InCorrPanel} and the bounding contacts are highlighted in the Outcrop View. This supports the geologists in identifying the region where they want to select cross bed measurements (\textbf{G5}). After confirmation, the selected cross beds are assigned to the selected stratum and a rose diagram appears next to the log. From this point on, selecting the stratum also selects the cross beds assigned to it. Changing this set by adding or removing cross beds is also reflected in the rose diagram (\textbf{G1}). We discussed approaches for the automatic selection of cross beds with our \rev{collaborators}, but in the end resorted to letting them select the measurements manually. After having created at least two logs, i.e. performing \textbf{T1a} twice, users can create correlations (\textbf{T2}).

\textbf{T2a Find Similar Strata:} Finding correlations means that geologists need to identify contacts that belong together, i.e. two contacts delineate the transition between the same strata in different logs. To achieve this, geologists need to be able to effectively compare outcrop characteristics encoded into a log, comprising the rock type of a stratum, its thickness, and the orientation distribution of the cross beds within it. To support this task, the InCorrPanel allows users to arbitrarily change the horizontal order of logs and compare any two logs via juxtaposition.

\textbf{T2b Connect Contacts to Create Correlations:} When suitable contacts are identified, the geologists select the respective lines, one per log. On confirmation, the selected contacts are then connected by curved lines in the color of the selected contacts. Each connection between two correlated logs is initially rendered in a dashed pattern to convey uncertainty and can be switched to a solid line style. If the correlation analysis is complete and the InCorrPanel contains all the findings the geologists want to present their results for scientific dissemination, for instance as an essential part of a geological publication, moving on to \textbf{T3}.

\textbf{T3 Level to Horizon and Export InCorrPanel:} To make logs more comparable in a published correlation panel, geologists typically level all logs to a common horizon (\textbf{G3}). This means to vertically align all logs to a common baseline correlation. When drawn in the same coordinate system, correlating contacts rarely occur on the same height, as exhibited by the red correlations in \autoref{fig:log}. Ideally there is a distinctive rock layer present in every log, which results in a correlation across the whole panel. Then this correlation can act as a vertical baseline. To achieve this, geologists select a correlation in the InCorrPanel and confirm to vertically align all logs to it. This results in straight connections instead of curved ones, as illustrated by \autoref{fig:incorrpanel}e. For final adaptions and editing, the content of the InCorrPanel can be exported as a scalable vector graphic readable by all common vector drawing applications (\textbf{G4}).

\subsection{\rev{Implementation}} \label{sec:impl}

InCorr is implemented in F\# as an extension of PRo3D~\cite{pro3dpage}, utilizing an ELM-style framework \cite{aardvark} for building scalable visual computing applications. \rev{Programming types and transformation functions follow the rules of domain driven design \cite{wlaschin2018domain} and align with the data model described in \autoref{sec:model}}. The InCorrPanel is created in the SVG format, allowing geologists to easily refine the exported visualization in a vector illustration software such as Adobe Illustrator. 
The 3D view, GUI elements, and the InCorrPanel are composed via a JavaScript code generator to run in a web browser.

\section{Evaluation} \label{sec:eval}

In this section we discuss the methods we used to evaluate InCorr. We verified InCorr by using it to recreate the correlation analysis of the HBDQ campaign. Then we validated the resulting InCorrPanel (\autoref{fig:incorrpanel}a) against the manually created illustrations shown in \autoref{fig:incorrpanel}b--d. Finally, we collected feedback from three geologists after performing tasks in the form of a \rev{design validation}.

\subsection{Use Case - Hanksville-Burpee Dinosaur Quarry} \label{sec:usecase}

\begin{figure}[]
 \centering
	\includegraphics [width=0.95\linewidth,trim=40 70 120 20,clip=true] {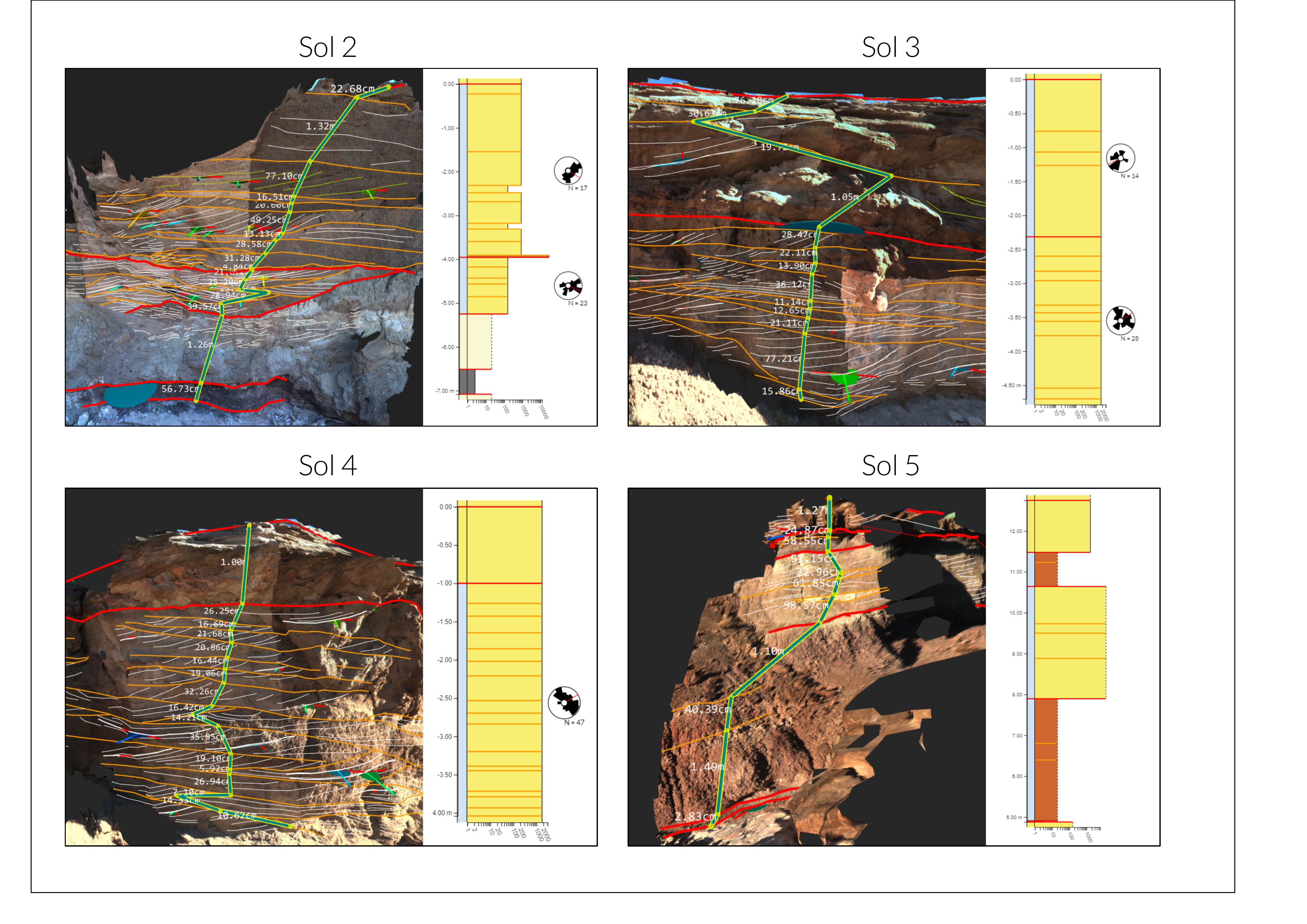}
 \caption{Outcrops Sol2--5 of the HBDQ dataset, each with a 3D log created with InCorr and the resulting 2D log from the InCorrPanel. \vspace{-0.4cm}}
 \label{fig:eval}
\end{figure}

We used InCorr to create \textbf{4} logs, one for each outcrop of the HBDQ dataset named Sol2--5, as shown in \autoref{fig:eval}. We used the logging tool to connect \textbf{62} contacts from the interpretation data, which amounted to \textbf{58} true thickness measurements. We assigned the rock types to the same number of strata and selected \textbf{18} strata to which we assigned a total of \textbf{129} cross bed measurements. We then added \textbf{2} correlations, which resulted in the correlation panel presented in \autoref{fig:incorrpanel}a. The whole process took approximately \textbf{1.5} hours. 
Carrying out this process in a non-assisted manner, using field data, involves manual measurement of the distances between contacts, conversion of apparent to true thickness, and manual plotting of rose diagrams. According to our collaborators, for the amount of data presented, this would take considerably longer than 1.5 hours, and though the time it would take would vary between workers, producing a correlation of equal precision would be on the order of a \textbf{day} to \textbf{several days} of work.

Our log representation matches the visual properties of the `simplified log illustrated manually' (\autoref{fig:incorrpanel}d) and so meets the minimal encoding capabilities, as proposed by our collaborators (\autoref{sec:singlelog}). We extended this version by two data-driven features, the secondary log and the rose diagrams. Our solution does not include the following styles present in the \rev{`log illustrated manually'} (\autoref{fig:incorrpanel}c): (1) curved, converging, or tilted contacts, (2) gradients, pattern fills, and rounded corners for strata, nor (3) glyphs to indicate additional rock properties.
Besides depending on the individual geologist's taste, most of these features do not directly encode information that is readily quantifiable in the data. We left these encodings to the manual workflow following the export of the InCorrPanel.
Nevertheless, it is interesting to incorporate styles that could be inferred from the data, such as curved or tilted contacts.
We could not recreate the thicknesses of one stratum in Sol~2 and one in Sol~5 (\autoref{fig:incorrpanel}f). The reason is either vertical exaggeration to make them visible or the true thickness is inferred from information not present in the data.
When we compare the \rev{`panel illustrated manually'} (\autoref{fig:incorrpanel}b) with the InCorrPanel as a whole, the main difference is that our illustration of the HBDQ campaign is flattened to the top correlation instead of the one connecting the two fossil layers.
This is because in the \rev{`panel illustrated manually'} the location of the fossil layer is assumed for Sol~3 and Sol~4 although it is actually buried in the physical outcrop and therefore the logs contain an empty region (\autoref{fig:incorrpanel}g).
We did not include this in InCorr, because it would require additional interactions and design iterations to make this data-driven and geologists can rarely make such assumptions on Mars.
We did not implement the visual connection of contacts with a line between non-adjacent logs. This would require a layout algorithm that avoids collisions with logs in between. At the same time, the resulting curved line should encode the course of the correlation, which can only be specified by the geologist.
But, a correlation can be created between arbitrary logs, and the visual connection will show when the respective logs are adjacent.

\vspace{-0.1cm}
\subsection{\rev{Design Validation}} 

\rev{We recruited three geologists to participate in our design validation and we will refer to them as P1, P2, and P3. P1 is a sedimentologist involved in design phases \textbf{(1)} and \textbf{(2)}, P2 is an engineering geologist from the field of tunnel constructions, and P3 is a planetary scientist with a geological background, none of which are co-authors of this paper.} We conducted one-to-one interviews limited to one hour, starting with background questions to their field and their expertise in DOM interpretation and logging. Then we asked them to perform the tasks of creating logs \textbf{T1} and creating correlations \textbf{T2} on two outcrops of the HBDQ dataset, closing with a short questionnaire and open feedback. Prior to the interview the participants received a five minute tutorial video, briefly explaining the interactions for achieving \textbf{T1} and \textbf{T2}. We decided to collect qualitative feedback on the relevance of the problem, the quality and ease of our solution, and its potential impact.

\head{time} All participants agreed, that logging and the correlation analysis of DOMs is a time-consuming process: \emph{`Often it is faster to draw the panel by hand on scales paper \emph{(opposed to using vector drawing software)'} (P3)}, and that InCorr can provide a significant speed up to logging and assigning cross beds: \emph{`Looking at this software I should be able to do it very quickly. About an hour instead of a week.' (P1)}.

They were very positive about the interactions in the 3D Outcrop View while creating a log and easily changing the succession of points: \emph{`Usually it is hard to get the scales correct' (P3)}, and \emph{`That works pretty well!' (P1)}. The linked highlighting was also a very welcome feature for relating logs and their spatial counterparts, but also when assigning cross beds: \emph{`The highlighting of a stratum's contacts is very intuitive. It feels visual and haptic.', `This is perfect for correlating spatially over large distances.' (P2).}

Each participant was missing different stylized features in the logs and correlations representation.
Asked if they could encounter the InCorrPanel `as is' in a geological publication they generally agreed, but they themselves would not use it directly without additional editing. 
Although P1 was content with post production \emph{`Really good that it is already in SVG, that's important.
Glyphs and styles don't need to happen in InCorr, I would export it to AI or Corel anyway'}. Still, he wished for more options on encoding different properties along the correlation lines. 
When two correlation lines end in one, it is interesting to show where the phase out happens. 
P2 wished for hatching patterns to fill strata and to be able to change the roundness of the rectangles to encode weathering effects. 
He also pointed out that he is unfamiliar with the quantification of the grain size in our rock types list. With our collaborators we agreed on displaying grain size categories as values of the Krumbein $\phi$ scale~\cite{Wilson1964}, but for broader acceptance we should also show the metric values and support other categorizations.
P3~immediately addressed the absence of glyphs and curved contacts.

When asked about the potential and implications of the tools provided through InCorr, each participant expressed their excitement, but for different reasons.
P1 plans to use the logging tool for a publication requiring extensive true thickness analysis.  
P2 is eager to try out logging and correlation analysis in lithological outcrops, i.e. tunnel faces created during tunnel excavation. P3 focused on the implications of a data-driven approach to correlation analysis, since in her current research she needs to reproduce and compare results from data that is only present in figures. On the suggestion of bundling the analysis and input data with a publication to be viewed with InCorr and PRo3D, she answered \emph{`Isn't this the whole point of a publication? To have other people interact with your data. I hope that is the future of publishing'}. 

\section{Discussion and Future Research} \label{sec:discussion}

In this section we summarize the key aspects of InCorr, relate them to the design goals we defined in \autoref{sec:goals}, and present a critical appraisal of our solution supported by the results from \autoref{sec:eval}, referring to participants of the \rev{design validation} where necessary.

We developed a 3D logging tool that allows geologists to create geological logs with little effort, especially when compared to current approaches (\textbf{G1}). All participants could create a 3D log in under a minute, while spending most of the time on considering which contacts to include. We had some usability issues when selecting contact lines or very thin strata in the InCorrPanel. To improve the selection process of thin lines, a list of potential candidates could be provided, combined with additional visual feedback, like a selection preview while hovering the lines or thin strata. Assigning cross beds to strata would greatly benefit from better selection tools provided by PRo3D, as for instance a line brush. Further, with the knowledge gained throughout this work we could try again to offer our collaborators an automatic pre-assignment of cross beds within a stratum that they can then intuitively edit.

Our data-driven approach of logging directly on the 3D contact lines ties logs and correlations to the outcrop interpretation data (\textbf{G2}). All participants recognized this as a significant improvement over current methods, where a log is rather a collection of height measurements than a 3D polyline. It would be interesting to investigate how to enrich the exported vector graphics file with meta data indicating its origin. According to \rev{the design feedback}, the InCorrPanel contains a sufficient set of visual encodings and customization to pass as a correlation panel in a geological publication (\textbf{G3}). In  \autoref{sec:eval} we could demonstrate, that InCorr encompasses all features to conduct a correlation analysis. Nevertheless, \textbf{G3} offers the largest room for improvement. The In\-Corr\-Panel establishes a baseline while some styling features are clearly left to vector drawing tools. It would be beneficial to explore the boundary of this separation, adding features that are supporting the analysis or should be reflected in the data.

The integration with PRo3D and the export of the InCorrPanel to a vector format allows InCorr to successfully bridge the gap between outcrop interpretation and creating regional geological models based on correlations (\textbf{G4}). We also verified this by reconstructing the correlation analysis of the HBDQ from existing outcrop interpretation data. On the other end, \autoref{fig:logvis} shows a modified log representation exported from the InCorrPanel. The highlighting of the selected log and of a stratum's bordering contacts allows users to establish a context between the 3D outcrop interpretation and the 2D log representation (\textbf{G5}). Geologists are very aware of the spatial context between the outcrops and logs of their analysis, due to the fact that they may spend weeks on interpreting the outcrops. None of the participants was familiar with the HBDQ data, so the contact highlights helped them to quickly orient themselves in the interpretation. We noticed that P3 subsequently tried to select strata in the Outcrop View, i.e. the region on the surface between two outcrops. Geometry for strata, which one could infer from contacts and the log, would benefit InCorr on multiple levels. It would be the basis for the assisted assignment of cross beds as mentioned before. One could directly assign rock types in the Outcrop View without switching to the panel, and further it could tie the actual DOM geometry and texture to the logging. P1 also entertained the idea of juxtaposing outcrops in 3D, aligning them along a geographic direction, analogous to the correlation panel. Such a transformation would dissolve longitude and latitude components of the data, while preserving elevation and orientation.

\head{Generalizability} InCorr was designed to support geologists in the geological analysis of Mars by translating static correlation panels to the interactive world. We could identify several other application areas for InCorr.
As suggested by P2, InCorr can be used to log the succession of strata in solid rock, exposed in the form of tunnel face outcrops. These are created and captured at regular intervals during tunnel excavation.
In large-scale tunnel construction projects it is common practice to maintain a reconnaissance tunnel, smaller and a few hundred meters ahead of the main tunnel.
Logging and correlation of the reconnaissance tunnel allows geologists to reconstruct the geological situation in the mountain and predict when the main tunnel will hit critical rock sections.
Outcrop-based logging and correlation panels are not used in this field of geology and we will pursue this idea in an ongoing research project. In the context of sedimentary geology and the exploration of Mars, we can adapt the InCorrPanel to display other measurement aggregations of strata than dip azimuths.
Our data model and interactions support to add, for instance, bedding thicknesses or grain sizes to strata, and show their distribution in histograms.
Finally, stratigraphic succession plays an essential role in archeological excavations and InCorr could be a valuable addition to the work of Traxler et al. \cite{traxler2008harris}.

Although correlation panels are a very domain-specific representation, we can generalize them as follows: a log connects discrete features to form intervals. The order is implicitly given by a metric and the intervals are nested through the rank of their features. Users connect two features in different logs through a correlation depending on the similarity of the two intervals the feature is part of. As an example, one could use the InCorrPanel to compare the lives of different people. Each person's life consists of events of different magnitude of change. These events are implicitly connected and sorted by their time of occurrence and form a log. Then one could correlate the life-changing events across multiple persons and investigate surrounding events or inspect similar phases of life and compare the events they begin and end with. From a geological perspective, it is obvious that correlation panels can be used to represent time, because they already do. The elevation of contacts and strata marks their occurrence in time, while the rank of a contact translates to the frequency of occurrence. For instance, the red contacts shown in \autoref{fig:outcrop} represent a change that may happen once every 100 to 1000 years, while the changes shown as orange contacts can happen once in a week or a month.

\vspace{-0.1cm}
\section{Conclusion}

With InCorr we present a design study of translating the static geological illustration of correlation panels for outcrop analysis to an interactive solution. It comprises a 3D logging tool and an interactive data-driven correlation panel, the InCorrPanel. We integrated both with an existing outcrop interpretation software and provided a vector graphics export to integrate with the established workflow of geologists. We demonstrate InCorr\rev{'}s functionality through a use case and validated our results against a manually created illustration based on identical outcrop interpretations. Further, we collected qualitative feedback from three geologists. The evaluation indicates that InCorr is a powerful and useful tool and can significantly improve geological analysis workflows through data propagation, and reducing time and effort of illustrating logs and correlation panels manually.

\acknowledgments{We wish to thank Sanjeev Gupta and Steve Banham from Imperial College of London for taking the time to share their knowledge about sedimentology. We also thank Marijn van Capelle for annotating the outcrop reconstructions provided by Joanneum Research and creating the manually illustrated correlation panel. This work was funded by ESA-PRODEX funding, supported by the Austrian Research Promotion Agency under ESA PEA Grants 4000105568 \& 4000117520 and by BMK, BMDW, Styria, SFG and Vienna Business Agency in the scope of COMET - Competence Centers for Excellent Technologies (854174) which is managed by FFG.}

\bibliographystyle{abbrv-doi}

\bibliography{visarbibliography}
\end{document}